\documentclass[aps,prx,reprint,nofootinbib,twocolumn,superscriptaddress,showpacs,showkeys,longbibliography]{revtex4-1}
\usepackage{eurosym}
\usepackage{amsmath,amssymb,amstext,bm}
\usepackage{ulem}
\usepackage[usenames,dvipsnames]{color}
\usepackage{graphicx}
\usepackage{braket}
\usepackage{natbib}
\usepackage{comment}
\usepackage{dcolumn}
\usepackage[english]{babel}
\usepackage{wasysym}

\usepackage[colorlinks,bookmarks=false,citecolor=blue,linkcolor=red,urlcolor=blue]{hyperref}

\begin{document}

\title{Floquet and anomalous Floquet Weyl semimetals}
\author{Yufei Zhu}
\affiliation{Department of Physics, Zhejiang Normal University, Jinhua 321004, China}
\author{Tao Qin}
\affiliation{Department of Physics, School of Physics and Materials Science, Anhui University,
Hefei, Anhui 230601, China}
\author{Xinxin Yang}
\affiliation{Department of Physics, Zhejiang Normal University, Jinhua 321004, China}
\author{Gao Xianlong}
\email{gaoxl@zjnu.edu.cn}
\affiliation{Department of Physics, Zhejiang Normal University, Jinhua 321004, China}
\author{Zhaoxin Liang }
\email{zhxliang@zjnu.edu.cn}
\affiliation{Department of Physics, Zhejiang Normal University, Jinhua 321004, China}
\date{\today}

\begin{abstract}
The periodic driving of a quantum system can enable new topological phases without analogs in static systems. This provides a route towards preparing nonequilibrium quantum phases rooted in the nonequilibrium nature by periodic driving engineering. Motivated by the ongoing considerable interest in topological semimetals, we are interested in the novel topological phases in the periodically driven topological semimetals without a static counterpart. We propose to design nonequilibrium topological semimetals in the regime of a weakly driving field where the spectrum width has the same magnitude as the driving frequency. We identify two types of nonequilibrium Weyl semimetals (i.e., Floquet and anomalous Floquet Weyl semimetals) that do not exhibit analogs in equilibrium. The proposed setup is shown to be experimentally feasible using the state-of-the-art techniques used to control ultracold atoms in optical lattices.
\end{abstract}

\maketitle

\section{Introduction}
At the heart of modern physics are the discovery and control of new phases of matter, highlighted by the foundational aspects of our understanding of periodic driving, which provides wholly new types of topological phases without analogs in equilibrium \cite{PhysRevLett.123.266803,Kitagawa2010,Jiang2011,Lindner2011,Rudner2013,Hu2016,Budich2017,Nathan2017,Martin2017,Yao2017,PhysRevResearch.1.022003,PhysRevB.100.085138,Gil2019,PhysRevLett.124.057001,PhysRevLett.124.216601}. A paradigmatic example is given by the periodically driven two-dimensional Dirac model \cite{Kitagawa2010,Lindner2011,Rudner2013,PhysRevA.91.043625,PhysRevB.90.115423,PhysRevLett.113.266801,PhysRevB.89.121401}, where robust chiral edge states can appear even though the Chern numbers of all bulk Floquet bands are zero. A system exhibiting this anomalous behavior has been realized recently using microwave photonic networks \cite{Hu2015}. Heretofore, such investigations of periodically driving a quantum system were largely restricted to static system that are topological insulators. Meanwhile, considerable effort has been devoted to the investigation of various intriguing phenomena associated with Weyl points such as the Fermi arc surface states \cite{Lv2015,Xu2015,PhysRevLett.116.096801,PhysRevLett.121.106402} and chiral anomaly~\cite{Zyuzin2012,Liu2013,Burkov2014,Parameswaran2014,Sun2015,Tan2019,Baireuther_2016,PhysRevLett.111.027201}.
Unlike topological insulators, whose gapless excitations are always at the sample boundary, topological semimetals host gapless fermions in the bulk. An outstanding challenge is to realize novel nonequilibrium topological phases in periodically driven topological semimetals without analogs in static systems. Another challenge is to extend the periodic driving from an insulator to semimetals with the emphasis on capturing the gapless nature of the energy spectra of the static counterpart
\cite{PhysRevLett.116.176401,PhysRevLett.123.066403,peri2018anomalous,YanYan2016,Huebener2017,Chen2018,Wang2014,Zhang2016}.
These considerations motivate the search for the novel scenarios of periodically driven topological Weyl semimetals.

\begin{figure}[!tbp]\centering
\centering
\includegraphics[width=.45\textwidth]{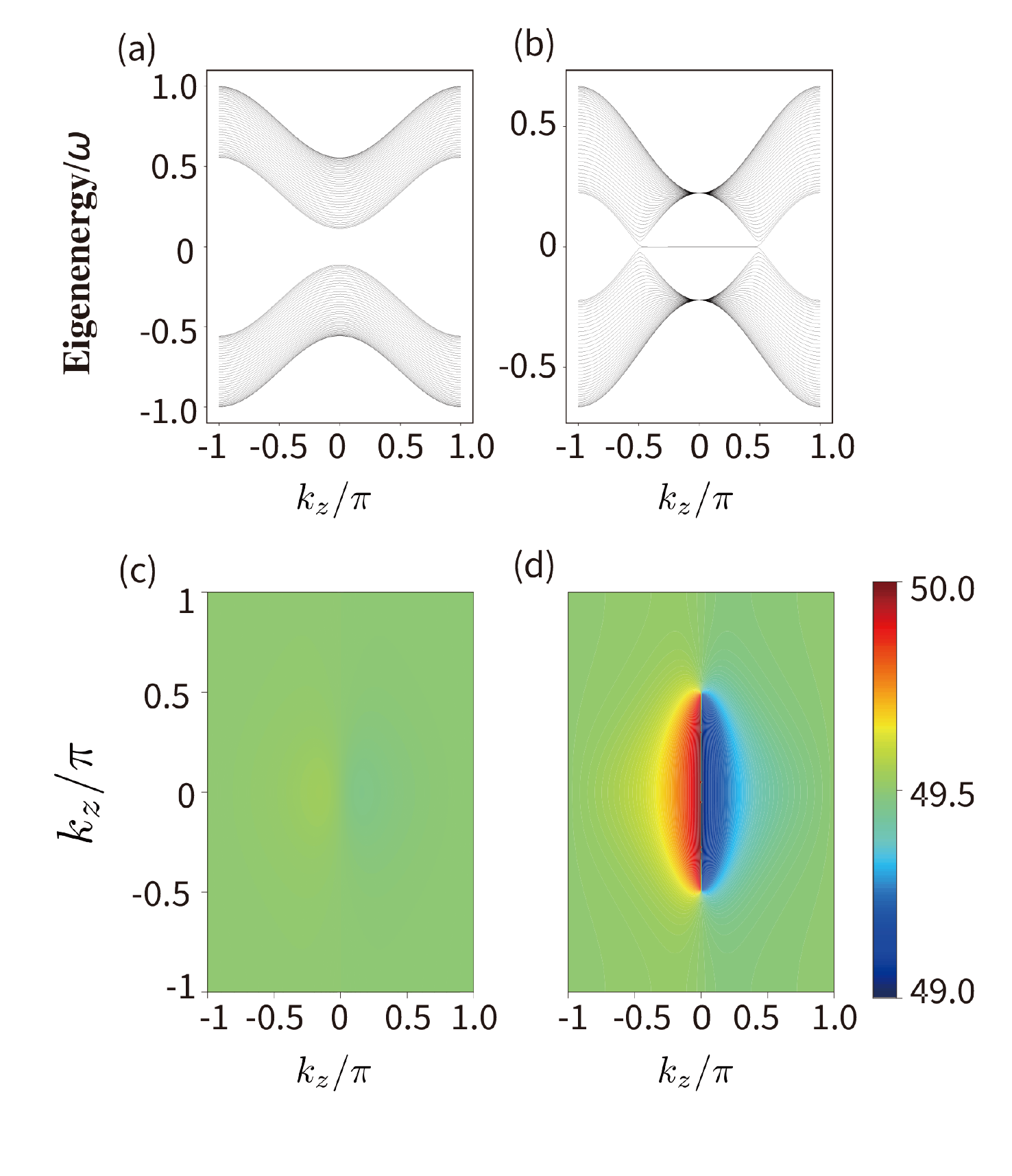}
\caption{Energy spectrum of the static Hamiltonian $H_{0}(k_x=0,y,k_z)$ in cylindrical geometry with edges along the $y$ direction and periodic in the $x$  and $z$  directions; $t_{0}=t_{s}=1.0$. We choose $t_0$ as the energy unit. The spectrum for $m=4.0t_0<{m_{c}}$ in (b) shows that $k_z=\pm\frac{\pi}{2}$ is the $z$ component for the position of Weyl points, while there are no Weyl points for $m=7.0t_0>{m_{c}}$ in (a). Correspondingly, the position of the hybrid Wannier center, directly related to the Chern number, is shown in (c) and (d) with the lattice being periodic in $y$ and $z$ directions and finite in the $x$ direction. (d) There is a shift in one lattice site if $k_{z}$ is between ${k_{c}=\pm{\frac{\pi}{2}}}$ with $m=7.0t_0$, with a shift smaller than one for $m=4.0t_0$.}\label{static}
\end{figure}

\begin{figure}[!tbp]\centering
\centering
\includegraphics[width=.50\textwidth]{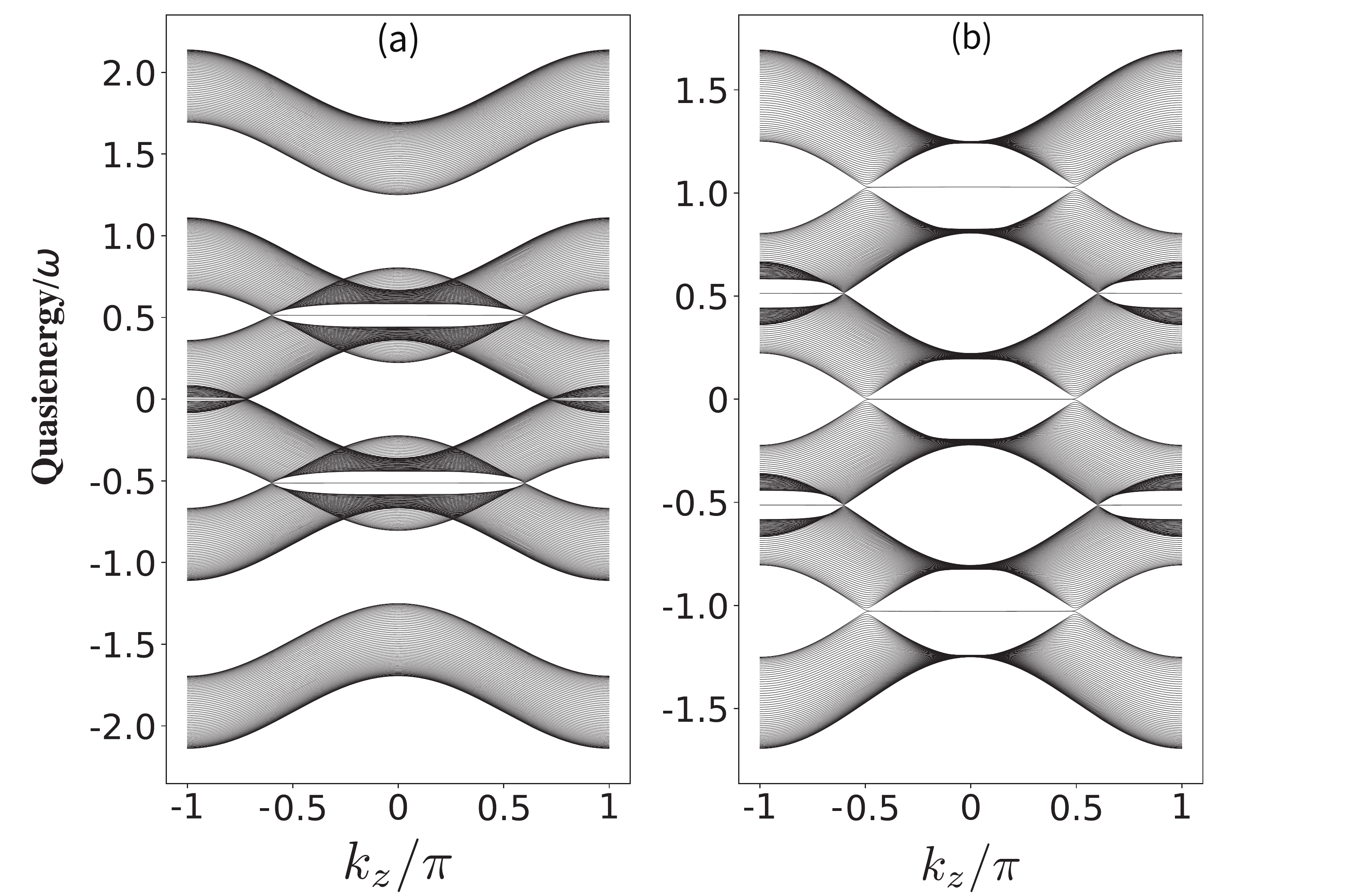}
\caption{Truncated Floquet Hamiltonian spectrum with the layer $k_{x}=0$ in the cylindrical geometry of a good quantum number $k_{z}$ and an open boundary in the $y$ direction.The horizontal axis denotes the crystal momentum $k_{z}$ and the vertical one denotes quasienergy in units of driving frequency $\omega$. The parameters are chosen as $\Delta=1.5$, $t_{0}=t_{s}=1.0$, $\omega=9$, and (a) $m=7.0t_0$ and (b) $m=4.0t_0$. According to the static topologically trivial case, the top and bottom bands should contribute zero Chern numbers. However, we localized edge modes that connected two Weyl points in different $m$-index bands.}
\label{Floquet}
\end{figure}

In this work we uncover and analyze two different types of nonequilibrium Weyl semimetals, i.e., Floquet and anomalous Floquet Weyl semimetals, which do not exhibit analogs in equilibrium. The former is referred to as the case of periodically driven systems, which are designed to be Floquet Weyl semimetals with the nontrivial topological phases, although the equilibrium counterpart is topologically trivial. In the latter case, we demonstrate the topological appearance in the stroboscopic dynamics of a periodically driven system.

We stress that the nonequilibrium Weyl semimetal induced in the intermediate-frequency limit as realized in this work relies on the unique topology of Floquet operators in the time-frequency domain and cannot appear in static systems. Our approach goes conceptually beyond the strength of the effective Hamiltonian based on high-frequency driving models by considering the full quasienergy spectrum without involving adiabatic projections, and is an alternative way to demonstrate nontrivial topological phases without static analogs. Furthermore, we propose that it is highly possible to realize our results in the context of ultracold atoms in optical lattices.
Hence our work is in contrast to those in Refs. \cite{Wang2014,Chen2018,Huebener2017}. There the focus is on how to use Floquet driving to induce a desired Floquet Hamiltonian describing Weyl semimetals whose topological characterizations are analogous to static systems. More specifically, Ref.~\cite{Huebener2017} demonstrated how femtosecond laser pulses with circularly polarized light can be used to induce, from a
prototypical three-dimensional Dirac material, a Floquet Hamiltonian that represents a Weyl semimetal, Dirac semimetal, or topological insulator. Reference.~\cite{Chen2018} presented a study of the circularly polarized light-induced Floquet states in type-II line-node
semimetals, while Ref.~ \cite{Wang2014} proposed that a Floquet Weyl semimetal can be induced in three-dimensional
topological insulators. Finally, Ref. \cite{Zhang_2016} presented a classification of different types of photoinduced Weyl points characterized by the Chern number. Instead, along the lines of Ref. \cite{Rudner2013}, our work focuses on topological phases unique to Floquet driving systems.

The structure of the paper is as follows. In Sec. \ref{Sec2} we present our theoretical model, based on which we solve for the topological phases without analogs in static systems. We identify two kinds of nonequilibrium quantum phases, i.e.,  Floquet and anomalous Floquet Weyl semimetals. In Sec. \ref{Sec3}, we present a comprehensive study of both nonequilibrium topological Weyl phases using numerical methods, providing an understanding in terms of the Chern number. In Sec. \ref{Sec4} we discuss how to experimentally prepare the Hamiltonian supporting the Weyl phases predicted in this work and conclude with a summary.

\section{Floquet and Anomalous Floquet Weyl Semimetals}\label{Sec2}

As the paradigmatic example of the two-band Hamiltonian on a square lattice hosting the topological Weyl semimetals \cite{Armitage2018} (WSMs) we consider its time-dependent lattice version
\begin{eqnarray}\label{Hk}
H(\textbf{k})&=&{\bf d}\left({\bf k}\right)\cdot \mathbf{\bm \sigma}+\Delta_0\sigma_z\cos\omega t,
\end{eqnarray}
where ${\bf d}=(2t_s\sin k_x,2t_s\sin k_y,m-2t_0(\cos k_x+\cos k_y+\cos k_z))$ and ${\bm \sigma}=(\sigma_x,\sigma_y,\sigma_z)$ are the $2\times 2$ Pauli matrices. The $t_s$ and $t_0$ of Hamiltonian (\ref{Hk}) denote the hopping amplitudes of spin-flipping hopping in the $x$-$y$ plane and spin-preserving hopping along all three dimensions, respectively, and $m$ represents the Zeeman field for tuning WSM states. The last term in Hamiltonian (\ref{Hk}) denotes the periodically driven term.

Without the periodically driven term, i.e., $\Delta_0=0$ in Hamiltonian (\ref{Hk}), two Weyl points (WPs) exist, which are located at the points $(0,0,\pm \arccos[(m-4t_0)/2t_0]$ for $2t_0<m<6t_0$. To characterize the topological properties of WSMs, treating $k_z$ as an effective parameter and reducing the original three-dimensional system to $k_z$-dependent effective two-dimensional subsystems is necessary. For a given $k_z\neq k_c$, the bulk bands are fully gapped, and the slice Chern number can be well defined with ${\bf n}=(d_x,d_y,d_z)/|{\bf d}({\bf k})|$ as $C\left(k_z\right)=1/\left(4\pi\right)\int d^2{\bf{k}} {\bf n}\cdot\partial_{k_x} {\bf n}\times \partial_{k_y} {\bf n}=1$ for  $2t_0<m<6t_0$ [see Figs. \ref{static} (b) and \ref{static} (d)] while $C\left(k_z\right)=0$ [see Figs. \ref{static} (a) and \ref{static} (c)]. The hybrid Wannier center is defined as $\left<n_x(k_y,k_z)\right>=\sum_{i_x}i_x\rho(i_x,k_y,k_z)/\sum_{i_x}\rho(i_x,k_y,k_z)$ where $i_x$ is the index for lattice sites and $\rho(i_x,k_y,k_z)$ is the occupied density of states. The Chern number is related to the hybrid Wannier center by $C=\delta\left<n_x(k_y,k_z)\right>$. Therefore, the WSM appears as a transitional state between a trivial insulator and a topological insulator as shown in Fig. \ref{static} (d), which can recover the previous results in Ref. \cite{Zhang2016} as expected.

With periodic driving [i.e., $\Delta_0\neq 0$ in the Hamiltonian (\ref{Hk})] the dynamics of the model system are governed by the time-periodic Hamiltonian of $H(t+T)=H(t)$. Our goal is to find the unique topological characteristics of periodically driven systems without analogs in equilibrium.

To illustrate the analogies to and differences from the static counterpart and in particular to further the construction of unique topological characteristics without the analogies of the static system, our strategy is to obtain the effective Hamiltonian of our model system within the rotating wave approximation by making the unitary transformation $U({\bf k},t)=P_++P_{-}e^{i\omega t}$. Here, $P_+$ and $P_-$ are the projectors on the upper and lower bands of the Hamiltonian (\ref{Hk}) with $\Delta_0=0$. Thus, the effective Hamiltonian is
\begin{equation}
H_{\text {eff}}=\left(\left|{\bf d}\right|-\frac{\omega}{2}-\frac{1}{2}\Delta_{0}\hat{d}_{z}\right){\bf d}\cdot{\bm \sigma}+\frac{1}{2}\Delta_{0}\sigma_{z}.\label{Heff}
\end{equation}
Specifically, the Hamiltonian (\ref{Heff}) can be simplified as $H_{\text {eff}}=\left(m-\omega/2\right)-2t_0(2+\cos k_z)\sigma_z$ when $k_x=k_y=0$. It is clear that with respect to the Hamiltonian (\ref{Hk}), the key physics underlying the Hamiltonian (\ref{Heff}) is the renormalization of the effective parameter $m_{\text{eff}}=m-\omega/2$, which induces the unique nonequilibrium topological phenomena [e.g., $2t_0<m_{\text{eff}}<6t_0$; see Figs. \ref{Floquet} (a)] from otherwise trivial static systems (e.g. $m>6t_0$). We note that a similar version of Floquet topological Weyl semimetals is given by Floquet topological insulators in Ref. \cite{Lindner2011}.

We are interested in the emergence of Floquet Weyl semimetals as counterparts of Floquet topological insulators~\cite{Lindner2011}. Starting with the topologically trivial phase ($m=8$), we study the effects of the periodic modulation of the Hamiltonian (\ref{Hk}) (the last term), which is supposed to create transitions between the valence and conduction bands at resonance and this typically opens gaps at the crossing points. The resulting bands exhibit nonzero Chern numbers, which correspond to Floquet Weyl semimetals, as shown in Fig.~\ref{Floquet} (a).

The above-mentioned intuitive understanding of the emergent Floquet Weyl semimetal is
limited to the analytical results of the high-frequency limit, i.e., the driving frequency $\omega$ is much larger than the other energy scale of the model system. In the most general case, the exact time evolution operator of the Hamiltonian can be computed as
\begin{eqnarray}\label{Ut}
U(\textbf{k},t)=\mathcal{T}\exp\left\{{-i\int_{0}^{t}H(\textbf{k},t^\prime)d{t^\prime}}\right\},
\end{eqnarray}
where $\mathcal{T}$ is the time-ordering operator. Note that $U$ is periodic in $k_x$, $k_y$, and $t$. Motivated by Ref. \cite{Rudner2013}, we can use $U$ to characterize the nonequilibrium
topological properties of the time-dependent Hamiltonian (\ref{Hk}) by defining a slice winding number
\begin{equation}
W_F(k_z)=\frac{1}{8\pi^2}\int dt d{\bf{k}}\text{Tr}\left(U^{-1}\partial_tU[U^{-1}\partial_{k_x}U,U^{-1}\partial_{k_y}U]\right).\label{Wind}
\end{equation}
On the basis of Eq. (\ref{Wind}), we emphasize that the underlying physical mechanism is that the micromotion that takes place within each driving period is crucial for the topological classification of periodically driven systems, which differs from the Chern number invariant in non-driven systems, which depend only on projectors onto the band of Floquet states.

Different types of nonequilibrium phases may emerge owing to the difference between the winding number and the Chern numbers of Floquet bands. By examining the gap centered at quasienergy $-\pi/T$, as shown in Fig. \ref{Floquet} (b), we expect to determine chiral edge modes spanning this gap on the basis of the effective Hamiltonian (\ref{Heff}). The further analytical calculation to the lowest order in $\Delta_0$ shows that the Chern number at quasienergy $-\pi/T$ is equal to zero. However, WPs between the bands of $l=0$ and $l=-1$ exist. We refer to this type of nonequilibrium Weyl semimetal as an anomalous Floquet Weyl semimetal, where the topology can only be explained by the winding number of Eq. (\ref{Wind}).

\section{Numerical results}\label{Sec3}

Above we developed the intuitive physical pictures of Floquet Weyl Semimetals and anomalous Floquet Weyl Semimetals.
Below we justify the existence of  Floquet Weyl semimetals and anomalous Floquet Weyl semimetals by solving the Hamiltonian (\ref{Hk}) with exact numerical methods.
\begin{figure*}[!tbp]\centering
\centering
\includegraphics[width=0.9\textwidth]{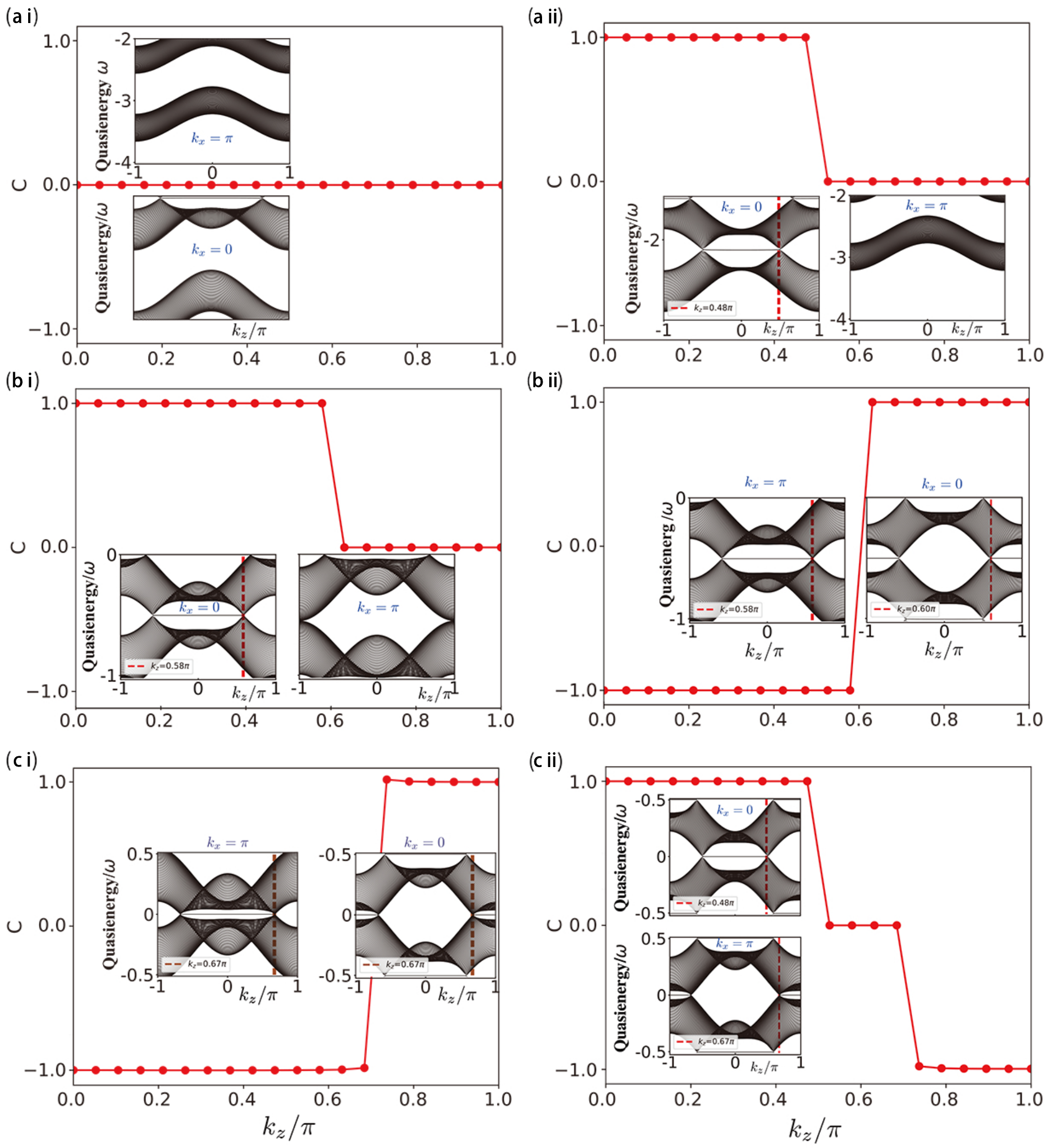}
\caption{Chern number for the lowest $LS/2-2$, $LS/2-1$ and $LS/2$ bands, and the corresponding quasienergy spectrum. $S=5$ in our calculation. A change in the Chern number reveals the position of the number of Weyl points. The parameters in left and right panels are the same as the corresponding ones in Fig.~\ref{Floquet}.}\label{Explain}
\end{figure*}

The general Floquet states corresponding to the Hamiltonian (\ref{Hk}) obey $[H(t)-i\partial_{t}]{\ket{\psi_{n}(t)}}=E_{n}\ket{\psi_{n}(t)}$, with $\ket{\psi_{n}(t)}=\ket{\psi_{n}(t+T)}$, and generate the quasienergy $ E_{n}$ (hereafter $\hbar=1$). We evolve Floquet states at any initial time $t=0$ by a time-evolution operator over one period. Thus, ${U}(T)\ket{\psi_{n}(0)}=e^{-iE_{n}T}\ket{\psi_{n}(0)}$ becomes an eigenvalue equation, where $e^{-iE_{n}T}$ is invariant under $E_{n}+l\omega$ $(l\in{\mathbb{N}})$. Hence, the quasienergy $E_{n}$ possesses periodicity and ${U}(T)=e^{-i{H}_{F}T}$.

The first Brillouin zone $-\omega\le{E_{n}\le\omega}$ in quasienergy contains all of the information that we are interested in and ${U}(T)$ becomes ${U}(\textbf{k},T)$, with $\textbf{k}$ the crystal momentum. More generally, the nondegenerate Floquet-Bloch time-evolution operator at $t$ is
\begin{eqnarray}\label{U}
U(\textbf{k},t)=\sum_{n=1}^{N}\ket{\epsilon_{n}(\textbf{k},t)}\bra{\epsilon_{n}(\textbf{k},t)}e^{-i\phi_{n}(\textbf{k},t)}
\end{eqnarray}
with the $n$th nondegenerate eigenstate $\ket{\epsilon_{n}(\textbf{k},t)}$ of $U(\textbf{k},t)$. Here $\phi_{n}(\textbf{k},t)$ is the pivot of quasienergy winding. In the meantime, we derive the Floquet Hamiltonian matrix as
\begin{eqnarray}\label{FlouqetH}
H^{l,l^{\prime}}_{\alpha,\alpha^{\prime}}(\textbf{k})&=&l\omega\delta_{\alpha,\alpha^{\prime}}\delta_{l,l^{\prime}}\nonumber\\
&&+\frac{1}{T}\int_{0}^{T}{dt}e^{-i(l-l^{\prime})\omega{t}}H_{\alpha,\alpha^{\prime}}(\textbf{k},t),
\end{eqnarray}
where $l$ and $l^{\prime}$ are the sector indices of quasienergy, and $\alpha$ and $\alpha^{\prime}$ are the dummy indices of the eigenbasis that depend on system size $L$. The extended Hilbert space of the effective Hamiltonian is $(L\times{S})^{2}$ (sector number $S$). The block diagonal of $H^{l,l^{\prime}}_{\alpha,\alpha^{\prime}}$ is dominated by the static Hamiltonian by adding $l\omega$. Other block off-diagonal elements of $H^{l,l^{\prime}}_{\alpha,\alpha^{\prime}}$ are the driving contribution.

First we consider how Floquet Weyl semimetals, as shown in Fig. \ref{Floquet}, can be generated when periodic driving is turned on.
Specifically, the static Hamiltonian is chosen to be topologically trivial with the parameters $m=7t_0$ and $t_s=t_0=1$. The band structure centered at quasienergy $0$ is shown in Fig. \ref{Floquet} (a). Then the periodic driving is turned on to the strength of $\Delta_0=1.5$. We can numerically solve the Floquet spectrum of the Hamiltonian (\ref{Hk}) for a cylindrical geometry with the periodic boundary conditions in the $z$ direction and open boundary conditions in the $y$ direction. By examining the band centered at quasienergy $0.5$, we see the appearance of the edge states, which suggests the existence of WPs. These Floquet Weyl semimetals can be explained in terms of the effective $m_{\text{eff}}=m-\omega/2$. Note that the static Hamiltonian (\ref{Hk}) is only topologically nontrivial in the parameter regime ($2t_0<m<6t_0$).  The initial value of $m=7t_0$ is out of
the topological parameter regime. However, the effective  $m_{\text{eff}}=m-\omega/2=2.5t_0$ is reduced by periodic driving, so the resulting effective Hamiltonian becomes topologically nontrivial.

Next we show how anomalous Floquet Weyl semimetals can be generated. Initially, the static Hamiltonian is chosen to be topologically nontrivial with two Weyl points with the parameters $m=4t_0$ and $t_s=t_0=1$ [as shown in Fig. \ref{static} (d)] or topological trivial without Weyl points [Fig.\ref{static} (c)]. Whether Weyl points in the band structure exist depends on the change in the value of $C(k_z)$, i.e., the Chern number of all bands below the gap. Following Ref. \cite{Lindner2011}, we can calculate $C(k_z)$ for the chosen gap. Then, the integer change in the Chern number indicates that the Weyl point can be detected with changing $k_z$. Indeed, as shown below, Fig. \ref{Floquet} (b) demonstrates the existence of anomalous Floquet Weyl semimetals.

Now we provide a deeper understanding in terms of the Chern number to show how the phenomenology
discussed in Eq. (\ref{Wind}) can arise in the microscopic lattice model without relying on adiabatic projections.
As pointed out in Ref. \cite {Rudner2013}, the key observation is that a straightforward
way to calculate the Chern numbers of bands for the truncated
Floquet Hamiltonian actually exists. Essentially, one can interpret the truncated Floquet
Hamiltonian as the static Hamiltonian of some new $N$-band system. With a change in the Chern number, we can demonstrate the position and number of the Weyl points.
We use Fig.~\ref{Explain} (c i) and (c ii) as examples to explain the main points. When $m=8.0t_0$, no Weyl points exist in the static Hamiltonian [Fig.~\ref{static} (c)]. Remarkably, with periodical driving, we see that in Fig.~\ref{Explain} (c i), four new Weyl points show up at $(\pi,0, k_z\approx\pm0.67\pi)$ and $(0,\pi, k_z\approx\pm0.67\pi)$. The jump in the Chern number is 2, which shows that two Weyl points appear simultaneously by tuning $k_z$. For $m=4.0t_0$, the Weyl points in Fig.~\ref{Explain} (c ii) at $(0,0,k_z\approx\pm0.48\pi)$ and $(\pi,\pi,k_z\approx\pm0.67\pi)$ are responsible for the jumps in the change in the Chern number.  The number is doubled compared to that of the static case. For the former, the mechanism is described in Eq.~\eqref{Heff}, where periodic driving can effectively change $m$. For the latter, the periodic driving can dramatically change the band structure to create new Weyl points. Therefore, periodic driving offers a way to manipulate Weyl points in the system that we explore. It can be a generic way to change the topological properties of quantum systems.

\section{Experimental realization and summary}\label{Sec4}

 Owing to the recent progress in the simulation of topological quantum matter with ultracold gases in optical lattices, the Hamiltonian (\ref{Hk}) can
be simulated experimentally with ultracold fermionic atoms in three-dimensional optical lattices. The spin degree of freedom can be encoded by two atomic internal states or
sublattices. Then the required spin-flipping hopping can be realized by synthetic spin-orbit coupling or a magnetic field in
the $xy$ plane; finally, the $\sigma_z$ term can be generated by similar external laser-atom dressing. The components that allow us to realize the Hamiltonian (\ref{Hk}) in optical lattices are well within the experimental reach with an extension. In a broader context, the tunable Weyl-semimetal bands can be simulated using superconducting quantum circuits. By driving the superconducting quantum circuits with microwave fields, Ref. \cite{Tan2019} mapped the momentum space of a lattice to the parameter space, which realized the Hamiltonian of the Weyl semimetal. At the same time, the topological winding numbers were further determined from the Berry curvature measurement. In summary, we hope that the predicted  Floquet and anomalous Floquet Weyl semimetals in this work can be observed in the abovementioned physical systems.

\section*{Acknowledgements}
We thank Ying Hu, Chao Gao, Wei Yi, and Biao Wu for inspiring discussion. This
work was supported by the key projects of the Natural Science
Foundation of China (Grants No. 11835011 and No. 11774316). T.Q. was
supported by a startup fund (Grant No. S020118002/069) from Anhui University.
\bibliography{ref_latest}

\end{document}